\documentclass[preprintnumbers,amsmath,amssymbm,preprint]{revtex4}
\usepackage{epsfig}
\usepackage{graphicx}

\begin{document}
\title{Infinitesimally thin static scalar shells surrounding charged Gauss-Bonnet black holes}
\author{Shahar Hod}
\affiliation{The Ruppin Academic Center, Emeq Hefer 40250, Israel}
\affiliation{ }
\affiliation{The Hadassah Institute, Jerusalem 91010, Israel}
\date{\today}

\begin{abstract}
\ \ \ We reveal the existence of a new form of spontaneously scalarized black-hole configurations. 
In particular, it is proved that Reissner-Nordstr\"om black holes in the highly charged regime $Q/M>(Q/M)_{\text{crit}}=\sqrt{21}/5$ can 
support {\it thin} matter shells that are made of massive scalar fields with a non-minimal 
coupling to the Gauss-Bonnet invariant of the curved spacetime. 
These static scalar shells, which become infinitesimally thin in the dimensionless large-mass $M\mu\gg1$ regime, hover a finite 
proper distance above the black-hole horizon [here $\{M,Q\}$ are respectively the mass and electric charge of the central supporting 
black hole, and $\mu$ is the proper mass of the supported scalar field]. 
In addition, we derive a remarkably compact analytical formula for the discrete resonance spectrum
$\{\eta(Q/M,M\mu;n)\}_{n=0}^{n=\infty}$ of the non-trivial coupling parameter 
which characterizes the bound-state charged-black-hole-thin-massive-scalar-shell cloudy configurations 
of the composed Einstein-Maxwell-scalar field theory.
\end{abstract}
\bigskip
\maketitle

\section{Introduction}

Wheeler's no-hair conjecture \cite{NHC,JDB} has asserted that black-hole solutions of the coupled Einstein-matter 
field equations should describe bald spacetimes in which spatially regular static scalar field configurations 
cannot be supported. 
In accord with this influential conjecture, various no-hair theorems \cite{Bek1,Sot2,Her1,Sot3,BekMay,Hod1} 
have ruled out the existence of scalarized black-hole spacetimes that are made of scalar fields with a minimal (and also with a 
non-trivial) coupling to the Ricci curvature scalar.    

Intriguingly, recent explorations \cite{Sot5,Sot1,GB1,GB2,ChunHer,SotN,Hodsg1,Hodsg2,Hodca,Done,Herrecnum,BeCo,Brih,Hernn} 
of the coupled Einstein-matter field equations have revealed the existence of hairy (scalarized) black-hole spacetimes 
in which the externally supported hair is non-trivially (non-minimally) coupled to the spatially-dependent 
Gauss-Bonnet curvature invariant ${\cal G}$ of the spacetime \cite{Notechar,Hersc1,Hersc2,Hodsc1,Hodsc2,Moh,Hodnrx}. 

This physically interesting phenomenon, known as black-hole spontaneous scalarization, is closely related to the fact that, 
in composed Einstein-Gauss-Bonnet-scalar field theories, the Klein-Gordon equation which determines 
the spatial behavior of the scalar field $\phi$ in the curved spacetime contains an effective mass term (a 
Gauss-Bonnet-scalar-field direct interaction term) of the linearized form $-\bar\eta\phi{\cal G}$. 
This spatially-dependent effective mass term leads, 
for large enough values of the non-trivial coupling parameter $\bar\eta$ [see Eq. (\ref{Eq28}) below], 
to the formation of a negative (attractive) black-hole-field binding potential well 
in the vicinity of the black-hole horizon. 

Recently, the physically important phenomenon of black-hole spontaneous scalarization has been explored, using 
numerical techniques, in the context of the charged black-hole solutions of the 
composed Einstein-Maxwell-Gauss-Bonnet-scalar field theory \cite{Brih,Hernn}. 
Interestingly, it has been demonstrated in \cite{Brih,Hernn} that the sharp 
boundary between bald black holes and hairy (scalarized) black-hole spacetimes is marked by the presence of 
charged Reissner-Nordstr\"om black holes that support spatially regular asymptotically flat 
bound-state configurations of the non-minimally coupled linearized scalar fields. 

The physical significance of the bound-state charged-black-hole-linearized-scalar-field cloudy configurations 
(the term scalar `clouds' is usually used in the physics literature to describe linearized 
field configurations that are supported by central black holes with spatially regular horizons \cite{Hodlit,Herlit}) stems 
from the fact that these composed configurations determine the 
charge-dependent critical existence-line $\bar\eta=\bar\eta(Q/M)$ \cite{NoteQb} of the non-trivial 
Einstein-Maxwell-Gauss-Bonnet-scalar field theory. 

The main goal of the present paper is to explore the physical and mathematical properties of non-minimally coupled 
linearized {\it massive} scalar field configurations (massive scalar clouds) that are supported by charged Reissner-Nordstr\"om 
black holes with spatially regular horizons. 
In particular, using analytical techniques, we shall reveal the physically intriguing fact that the addition of a {\it mass} term to the 
supported non-minimally coupled scalar fields [see the action (\ref{Eq1}), which characterizes the composed 
Einstein-Maxwell-massive-scalar field theory] allows the existence of infinitesimally thin static 
scalar shells that hover a {\it finite} proper distance above the horizons of 
highly charged Gauss-Bonnet black holes. 

In addition, below we shall derive a remarkably compact analytical resonance formula that describes, in 
the large-mass $M\mu\gg1$ regime, 
the functional dependence $\bar\eta=\bar\eta(Q/M,M\mu)$ of the critical existence line, which 
characterizes the composed charged-black-hole-thin-massive-scalar-shell bound-state cloudy configurations, on the 
electric charge $Q$ of the central supporting black hole and on the dimensionless proper mass $M\mu$ \cite{Notemuu} 
of the non-minimally coupled scalar field.  

\section{Description of the system}

We study the physical and mathematical properties of `cloudy' black-hole configurations which are made of central 
charged Reissner-Nordstr\"om black holes that support spatially regular bound-state static configurations of linearized 
{\it massive} scalar fields. 
The composed Einstein-Maxwell-Gauss-Bonnet-nonminimally-coupled-massive-scalar field 
theory is characterized by the action \cite{Brih,Hernn,Noteun}
\begin{equation}\label{Eq1}
S=\int
d^4x\sqrt{-g}\Big[{1\over4}R-{1\over4}F_{\alpha\beta}F^{\alpha\beta}-{1\over2}\nabla_{\alpha}\phi\nabla^{\alpha}\phi
-{1\over2}\mu^2\phi^2+f(\phi){\cal
G}\Big]\  ,
\end{equation}
where $\mu$ is the mass of the non-minimally coupled scalar field. 
As we shall explicitly prove below, the supported matter configurations owe their existence to the presence, in the 
action (\ref{Eq1}), of a direct (non-minimal) coupling between the massive scalar field $\phi$ and 
the Gauss-Bonnet invariant 
\begin{equation}\label{Eq2}
{\cal G}\equiv R_{\mu\nu\rho\sigma}R^{\mu\nu\rho\sigma}-4R_{\mu\nu}R^{\mu\nu}+R^2\
\end{equation}
that characterizes the curved black-hole spacetime. 

As discussed in \cite{GB1,GB2,Hersc1}, the leading-order functional behavior 
\begin{equation}\label{Eq3}
f(\phi)={1\over2}\eta\phi^2\
\end{equation}
of the scalar coupling function guarantees that the bald Reissner-Nordstr\"om black-hole 
spacetime is a valid solution of the composed Einstein-Maxwell-scalar field equations 
in the weak-field $\phi\to0$ limit \cite{Noteunir}. 
The strength of the non-minimal coupling between the supported massive scalar field 
and the Gauss-Bonnet curvature invariant (\ref{Eq2}) is controlled by 
the physical parameter $\eta$ \cite{Noteetaa}. 

The curved line element \cite{ThWe,Chan,Notebl}
\begin{eqnarray}\label{Eq4}
ds^2=-h(r)dt^2+{{1}\over{h(r)}}dr^2+r^2d\theta^2+r^2\sin^2\theta d\phi^2\
\end{eqnarray}
with
\begin{equation}\label{Eq5}
h(r)\equiv 1-{{2M}\over{r}}+{{Q^2}\over{r^2}}\
\end{equation}
characterizes the supporting Reissner-Nordstr\"om black hole of mass $M$ 
and electric charge $Q$. The roots 
\begin{equation}\label{Eq6}
r_{\pm}=M\pm(M^2-Q^2)^{1/2}\
\end{equation}
of the metric function (\ref{Eq5}) determine the radii of the (outer and inner) black-hole horizons. 

The action (\ref{Eq1}) of the composed Einstein-Maxwell-massive-scalar field theory yields 
the Klein-Gordon differential equation \cite{Hernn}
\begin{equation}\label{Eq7}
\nabla^\nu\nabla_{\nu}\phi=\mu^2_{\text{eff}}\phi\
\end{equation}
for the eigenfunctions of the supported scalar field configurations, where the effective mass term 
\begin{equation}\label{Eq8}
\mu^2_{\text{eff}}(r;M,Q)=\mu^2-\eta{\cal G}\  ,
\end{equation}
which depends on the Gauss-Bonnet curvature invariant
\begin{equation}\label{Eq9}
{\cal G}_{\text{RN}}(r;M,Q)={{8}\over{r^8}}\big(6M^2r^2-12MQ^2r+5Q^4\big)\
\end{equation}
of the Reissner-Nordstr\"om black-hole spacetime (\ref{Eq4}), 
reflects the non-trivial massive-scalar-field-Gauss-Bonnet coupling in the composed field theory (\ref{Eq1}). 

Intriguingly, one finds that, depending on the relative magnitudes of the physical 
parameters $\{\eta,\mu\}$ of the composed field theory (\ref{Eq1}), 
the radially-dependent effective mass term (\ref{Eq8}) may become 
negative in the vicinity of the black-hole outer horizon. 
Below we shall use analytical techniques in order to prove that this property 
of the effective scalar-field-Gauss-Bonnet mass term (\ref{Eq8}) may allow the existence of infinitesimally {\it thin} non-minimally 
coupled massive scalar shells (thin massive scalar clouds) that hover a {\it finite} proper distance above the horizons of 
highly charged [see Eq. (\ref{Eq23}) below] Reissner-Nordstr\"om black holes. 

Substituting the functional decomposition 
\begin{equation}\label{Eq10}
\phi(r,\theta,\varphi)=\sum_{lm} R_{lm}(r)Y_{lm}(\theta,\varphi)\
\end{equation}
for the static non-minimally coupled massive scalar field into Eq. (\ref{Eq7}) 
[here $Y_{lm}(\theta,\varphi)$ with $l\geq|m|$ are the familiar spherical harmonic functions] 
and using the curved black-hole line element (\ref{Eq4}), one obtains 
the radial differential equation \cite{Hernn,Noteomt}
\begin{eqnarray}\label{Eq11}
{{d}\over{dr}}\Big[r^2h(r){{dR}\over{dr}}\Big]-[\mu^2 r^2+l(l+1)]R+
\eta\Big({{48M^2}\over{r^4}}-{{96MQ^2}\over{r^5}}+{{40Q^4}\over{r^6}}\Big)R=0\  ,
\end{eqnarray}
which determines the spatial behavior of the supported massive scalar clouds in the 
curved black-hole spacetime (\ref{Eq4}). 

The radial differential equation (\ref{Eq11}), supplemented by the physically
motivated boundary conditions of spatially regular (bounded) 
functional behavior of the scalar field at the black-hole outer horizon \cite{GB1,GB2,Macnn}, 
\begin{equation}\label{Eq12}
\psi(r=r_{\text{H}})<\infty\  ,
\end{equation}
and an asymptotic exponential decay of the massive scalar eigenfunction 
at spatial infinity \cite{GB1,GB2,Macnn},
\begin{equation}\label{Eq13}
\psi(r\to\infty)\sim r^{-1}e^{-\mu r}\to 0\ ,
\end{equation}
determine the discrete resonance spectrum
$\{M^{-2}\eta(M,Q,\mu;n)\}_{n=0}^{n=\infty}$ of the dimensionless coupling parameter that 
characterizes the bound-state 
Reissner-Nordstr\"om-black-hole-nonminimally-coupled-massive-scalar-field cloudy configurations 
of the composed Einstein-Maxwell-Gauss-Bonnet-scalar field theory (\ref{Eq1}). In particular, the critical
existence-line of the physical system, which marks the boundary between bald Reissner-Nordstr\"om black holes 
and hairy (scalarized) black-hole solutions, is determined by the fundamental ($n=0$) 
resonant mode $\eta_0=\eta_0(M,Q,\mu)$ [see Eq. (\ref{Eq48}) below]. 

\section{The discrete resonance spectrum of the 
composed black-hole-nonminimally-coupled-linearized-massive-scalar-field configurations}

In the present section we shall prove the intriguing existence, 
in the composed Einstein-Maxwell-Gauss-Bonnet-massive-scalar field theory (\ref{Eq1}), of infinitesimally {\it thin} scalar shells 
with positive ($\eta>0$) values of the non-minimal Gauss-Bonnet-scalar-field coupling parameter that 
are supported in highly-charged black-hole spacetimes 
a {\it finite} proper distance above the black-hole horizons. 
In addition, we shall use analytical techniques in the dimensionless large-mass regime
\begin{equation}\label{Eq14}
M\mu\gg1\
\end{equation}
in order to determine the discrete resonance spectrum
$\{\eta(M,Q,\mu;n)\}_{n=0}^{n=\infty}$ of the 
composed Reissner-Nordstr\"om-black-hole-nonminimally-coupled-massive-scalar-field bound-state 
cloudy configurations. 

In particular, we shall explicitly prove that the composed 
Reissner-Nordstr\"om-black-hole-nonminimally-coupled-massive-scalar-field system is amenable to an 
analytical treatment in the dimensionless large-mass regime (\ref{Eq14}), which 
corresponds to the dimensionless large-coupling regime
\begin{equation}\label{Eq15}
\bar\eta\equiv {{\eta}\over{M^2}}\gg1\  .
\end{equation}

To this end, it is convenient to define the radial scalar eigenfunction
\begin{equation}\label{Eq16}
\psi\equiv rR\  ,
\end{equation}
in terms of which the radial equation (\ref{Eq11}) can be expressed in the mathematically compact form 
\begin{equation}\label{Eq17}
{{d^2\psi}\over{dy^2}}-V\psi=0\  ,
\end{equation}
where the differential relation \cite{Noteppmm}
\begin{equation}\label{Eq18}
dy={{dr}\over{h(r)}}\
\end{equation}
determines the new radial coordinate $y(r)$. 
The effective potential in the Schr\"odinger-like radial differential equation (\ref{Eq17}), 
which characterizes the composed black-hole-massive-scalar-field system, is given by the (rather cumbersome) 
functional expression 
\begin{eqnarray}\label{Eq19}
V(r;M,Q,\mu,l,\bar\eta)=\Big(1-{{2M}\over{r}}+{{Q^2}\over{r^2}}\Big)
\Big[\mu^2+{{l(l+1)}\over{r^2}}+{{2M}\over{r^3}}-{{2Q^2}\over{r^4}}
-\bar\eta\cdot V_{\text{GB}}\Big]\  .
\end{eqnarray}
The presence of the Gauss-Bonnet term 
\begin{equation}\label{Eq20}
V_{\text{GB}}(r;M,Q)={{48M^4}\over{r^6}}-{{96M^3Q^2}\over{r^7}}+{{40M^2Q^4}\over{r^8}}\
\end{equation}
in the effective interaction potential (\ref{Eq19}) is a direct consequence of the non-trivial (non-minimal) 
coupling between the Gauss-Bonnet curvature invariant (\ref{Eq9}) of the charged black-hole spacetime (\ref{Eq4}) and 
the supported massive scalar field. 

We shall henceforth consider composed black-hole-massive-scalar-field cloudy configurations in 
the dimensionless large-mass regime (\ref{Eq14}) [or equivalently, in the dimensionless large-coupling regime (\ref{Eq15})], 
in which case the effective black-hole-massive-field interaction potential (\ref{Eq19}) can be written 
in the form \cite{Notenhm}
\begin{equation}\label{Eq21}
V(r;M,Q,\mu,\bar\eta)=h(r)\Big[\mu^2-
\bar\eta\cdot\Big({{48M^4}\over{r^6}}-{{96M^3Q^2}\over{r^7}}+{{40M^2Q^4}\over{r^8}}\Big)\Big]\cdot\{1+O[(M\mu)^{-2}]\}\  .
\end{equation}

The Gauss-Bonnet term (\ref{Eq20}) has a peak whose charge-dependent radius is 
given by the simple functional relation
\begin{equation}\label{Eq22}
r_{\text{peak}}(M,Q)={{5Q^2}\over{3M}}\  .
\end{equation}
Interestingly, one finds that, in the dimensionless charge-to-mass ratio regime
\begin{equation}\label{Eq23}
{{Q}\over{M}}\geq \Big({{Q}\over{M}}\Big)_{\text{crit}}={{\sqrt{21}}\over{5}}\  ,
\end{equation}
the radial peak (\ref{Eq22}) is located {\it outside} the black-hole outer horizon [that is, $r_{\text{peak}}\geq r_+$ in the 
regime (\ref{Eq23})]. 
Below we shall prove that this intriguing physical property 
of the non-trivial Gauss-Bonnet term (\ref{Eq20}) allows the existence of infinitesimally thin massive scalar shells 
that hover a finite proper distance above the outer horizons of central supporting Reissner-Nordstr\"om black holes in the 
highly charged regime (\ref{Eq23}) \cite{Notealso,Notensz}.

Substituting (\ref{Eq22}) into Eq. (\ref{Eq20}), one finds the functional relations \cite{NoteQQc}
\begin{equation}\label{Eq24}
\text{max}_r\big\{V_{\text{GB}}(r\geq r_+;M,Q)\big\}=
\begin{cases}
{{17496M^{10}}\over{78125Q^{12}}} &\ \ \ \ \text{for}\ \  {{Q}/{M}}\geq ({{Q}/{M}})_{\text{crit}}\  \\
{{8[5M^2Q^4+12M^3(M^2-Q^2)^{3/2}-18M^4Q^2+12M^6]}\over{[M+(M^2-Q^2)^{1/2}]^8}} &\ \ \ \ 
\text{for}\ \  {{Q}/{M}}\leq ({{Q}/{M}})_{\text{crit}}\ \  
\end{cases}
\end{equation}
for the maximal value of the non-trivial Gauss-Bonnet term (\ref{Eq20}) in the exterior regions of charged 
Reissner-Nordstr\"om black holes. 

\subsection{Upper bound on the proper masses of non-minimally coupled scalar clouds} 

In the present subsection we shall derive a charge-dependent upper bound on the allowed proper masses of the 
non-minimally coupled scalar fields that can be supported by the central charged 
Reissner-Nordstr\"om black holes. To this end, we point out that the presence of a binding (attractive) potential well outside the 
black-hole outer horizon provides a necessary condition for the existence of static bound-state 
scalar field configurations (scalar clouds) that are supported in the curved black-hole spacetime. 

In particular, the requirement 
\begin{equation}\label{Eq25}
V(r_{t_-}\leq r\leq r_{t_+})\leq0\ \ \ \ \text{with}\ \ \ \ r_{t_-}\geq r_+\
\end{equation}
[here $\{r_{t_-},r_{t_+}\}$ with $r_{t_-}\geq r_+$ are the characteristic classical turning points of the 
effective curvature potential (\ref{Eq19})] yields the series of inequalities \cite{NoteMQ}
\begin{equation}\label{Eq26}
\mu^2-\bar\eta\cdot \text{max}_r\big\{V_{\text{GB}}(r)\big\}\leq\mu^2-\bar\eta\cdot V_{\text{GB}}(r)\leq\mu^2+{{l(l+1)}\over{r^2}}+{{2M}\over{r^3}}-{{2Q^2}\over{r^4}}-\bar\eta\cdot V_{\text{GB}}(r)\leq0\ \ \ \ \text{for}\ \ \ \ {\bar\eta>0}\  .
\end{equation}
Taking cognizance of Eqs. (\ref{Eq24}) and (\ref{Eq26}), one finds the charge-dependent upper bound 
\begin{equation}\label{Eq27}
M\mu\leq\sqrt{\bar\eta}\cdot 
\begin{cases}
{{54\sqrt{30}}\over{625}}\cdot\big({{M}\over{Q}}\big)^6 &\ \ \ \ \text{for}\ \ \ \ {{Q}\over{M}}\geq {{\sqrt{21}}\over{5}}\  \\
{{\sqrt{8[5M^4Q^4+12M^5(M^2-Q^2)^{3/2}-18M^6Q^2+12M^8]}}\over{[M+(M^2-Q^2)^{1/2}]^4}} &\ \ \ \ \text{for}\ \ \ \ {{Q}\over{M}}\leq {{\sqrt{21}}\over{5}}\ \  
\end{cases}
\end{equation}
on the allowed proper masses of the supported non-minimally coupled scalar fields. 

Interestingly, the upper bound (\ref{Eq27}) can be expressed as the lower bound 
\begin{equation}\label{Eq28}
\bar\eta\geq (M\mu)^2\cdot
\begin{cases}
{{78125}\over{17496}}\cdot\big({{Q}\over{M}}\big)^{12} &\ \ \ \ \text{for}\ \ \ \ {{Q}\over{M}}\geq {{\sqrt{21}}\over{5}}\  \\
{{[M+(M^2-Q^2)^{1/2}]^8}}\over{8[5M^4Q^4+12M^5(M^2-Q^2)^{3/2}-18M^6Q^2+12M^8]}&\ \ \ \ \text{for}\ \ \ \ {{Q}\over{M}}\leq {{\sqrt{21}}\over{5}}\ \  
\end{cases}
\end{equation}
on the dimensionless value of the non-minimal coupling parameter which characterizes the composed 
charged-black-hole-nonminimally-coupled-linearized-massive-scalar-field bound-state configurations. 

\subsection{The resonance spectrum of the composed black-hole-non-minimally-coupled-massive-scalar-field cloudy configurations}

In the present subsection we shall focus our attention on the dimensionless super-critical charge regime 
[see Eq. (\ref{Eq23})] 
\begin{equation}\label{Eq29}
{{Q}\over{M}}\geq {{\sqrt{21}}\over{5}}\
\end{equation}
which, as we shall explicitly prove below, is characterized by the presence 
of arbitrarily thin scalar shells that hover a finite proper distance above the central charged black holes [see 
Eq. (\ref{Eq43}) below]. 

Interestingly, we shall prove that the discrete resonance spectrum $\{\bar\eta(M,Q,\mu;n)\}_{n=0}^{n=\infty}$ of the 
dimensionless physical parameter ${\bar\eta}$, which characterizes the non-minimally coupled Einstein-Maxwell-massive-scalar 
field theory (\ref{Eq1}), can be determined analytically in the eikonal large-mass regime (\ref{Eq14}) [which 
corresponds to the large-coupling regime (\ref{Eq15}), see Eq. (\ref{Eq28})]. 
In particular, the radial differential equation of the supported non-minimally coupled massive scalar fields with the familiar 
Schr\"odinger-like form (\ref{Eq17}) is characterized by the well known 
second-order WKB quantization condition \cite{WKB1,WKB2,WKB3}
\begin{equation}\label{Eq30}
\int_{y_{t_-}}^{y_{t_+}}dy\sqrt{-V(y;M,Q,M\mu,\bar\eta)}=\big(n+{1\over2}\big)\cdot\pi\
\ \ \ ; \ \ \ \ n=0,1,2,...\  ,
\end{equation}
where the classical turning points of the effective black-hole-field binding potential (\ref{Eq21}) 
[with $V(y_{t_-})=V(y_{t_+})=0$] determine the integration limits in (\ref{Eq30}) and $n\in\{0,1,2,...\}$ is the discrete 
resonance parameter of the physical system. 
The WKB resonance condition (\ref{Eq30}) can be expressed, using the differential relation (\ref{Eq18}), 
in the mathematically more convenient form
\begin{equation}\label{Eq31}
\int_{r_{t_-}}^{r_{t_+}}dr \sqrt{-{{V(r;M,Q,M\mu,\bar\eta)}\over{[h(r)]^2}}}=\big(n+{1\over2}\big)\cdot\pi\
\ \ \ ; \ \ \ \ n=0,1,2,...\  .
\end{equation}

Using the dimensionless relations [see Eqs. (\ref{Eq22}) and (\ref{Eq27})]
\begin{equation}\label{Eq32}
M\mu=\sqrt{\bar\eta}\cdot {{54\sqrt{30}}\over{625}}\cdot\Big({{M}\over{Q}}\Big)^6\cdot(1-\epsilon)\ \ \ \ ; \ \ \ \ \epsilon\geq0\
\end{equation}
and
\begin{equation}\label{Eq33}
r\equiv r_{\text{peak}}\cdot(1+x)\  ,
\end{equation}
which define the auxiliary physical variables $\{\epsilon,x\}$, 
one can write the effective radial potential (\ref{Eq21}) of the composed 
black-hole-non-minimally-coupled-massive-scalar-field cloudy configurations in the dimensionless form \cite{Notexep}
\begin{equation}\label{Eq34}
M^2{{V(r)}\over{[h(r)]^2}}=\bar\eta\cdot{{34992Q^2}\over{3125(25Q^2-21M^2)}}\Big({{M}\over{Q}}\Big)^{12}
\big(-\epsilon+9\cdot x^2\big)
\cdot\Big\{1+O\Big[{{x}\over{25(Q/M)^2-21}},{{\epsilon}\over{[25(Q/M)^2-21]^2}}\Big]\Big\}\  ,
\end{equation}
where we have used here the functional relation [see Eqs. (\ref{Eq5}), (\ref{Eq22}), and (\ref{Eq33})] 
\begin{equation}\label{Eq35}
h(r)={{25Q^2-21M^2}\over{25Q^2}}
\cdot\Big\{1+O\Big[{{x}\over{25(Q/M)^2-21}}\Big]\Big\}\  .
\end{equation}

Substituting Eqs. (\ref{Eq22}), (\ref{Eq33}), and (\ref{Eq34}) into Eq. (\ref{Eq31}) and defining
\begin{equation}\label{Eq36}
z={{3}\over{\sqrt{\epsilon}}}\cdot x\  ,
\end{equation}
one obtains the characteristic WKB resonance condition
\begin{equation}\label{Eq37}
\epsilon\cdot{{12\sqrt{15}}\over{25}}\Big({{M}\over{Q}}\Big)^{4}\sqrt{{{\bar\eta Q^2}\over{25Q^2-21M^2}}}
\int_{-1}^{1}dz \sqrt{1-z^2}=\big(n+{1\over2})\cdot\pi\ \ \ \ ; \ \ \
\ n=0,1,2,...\    
\end{equation}
for the composed black-hole-massive-field system, which yields the 
discrete resonance spectrum \cite{Noteintegral}
\begin{equation}\label{Eq38}
\epsilon(Q/M,{\bar\eta})={{1}\over{\sqrt{\bar\eta}}}\cdot{{25}\over{6\sqrt{15}}}\Big({{Q}\over{M}}\Big)^{4}\sqrt{{{25Q^2-21M^2}\over{Q^2}}}
\cdot\big(n+{1\over2})\ \ \ \ ; \ \ \ \ n=0,1,2,...\  .
\end{equation}
Interestingly, and most importantly for our analysis, one finds from (\ref{Eq38}) the characteristic relation 
$\epsilon\ll1$ in the large-coupling $\bar\eta\gg1$ regime. 

Substituting the analytically derived relation (\ref{Eq38}) into (\ref{Eq32}), one obtains the 
large-mass (large-coupling) resonance formula
\begin{equation}\label{Eq39}
\sqrt{\bar\eta(Q/M,M\mu;n)}={{625}\over{54\sqrt{30}}}\Big({{Q}\over{M}}\Big)^{6}\cdot M\mu+
{{25}\over{6\sqrt{15}}}\Big({{Q}\over{M}}\Big)^{4}\sqrt{{{25Q^2-21M^2}\over{Q^2}}}
\cdot\big(n+{1\over2})\ \ \ \ ; \ \ \ \ n=0,1,2,...\
\end{equation}
for the composed Reissner-Nordstr\"om-black-hole-nonminimally-coupled-massive-scalar-field cloudy configurations.

\section{The effective radial widths of the supported scalar clouds} 

In the present section we shall determine the effective widths of supported 
massive scalar field configurations in the charged Reissner-Nordstr\"om black-hole spacetime (\ref{Eq4}). 
In particular, we shall explicitly prove that, in the dimensionless 
large-mass $M\mu\gg1$ regime, the supported scalar clouds can be made arbitrarily thin. 

The effective widths of the supported matter configurations in the charged black-hole spacetime 
are determined by the classically allowed radial region 
\begin{equation}\label{Eq40}
\Delta r(Q/M,M\mu)\equiv r_{t_+}-r_{t_-}\
\end{equation}
of the effective binding potential (\ref{Eq21}), which characterizes the composed black-hole-massive-scalar-field system (\ref{Eq1}). 
Taking cognizance of Eqs. (\ref{Eq22}), (\ref{Eq33}), and (\ref{Eq36}) with $\Delta z=1-(-1)=2$ [see Eq. (\ref{Eq37})], 
one finds the remarkably simple functional expression
\begin{equation}\label{Eq41}
{{\Delta r(Q/M,M\mu)}\over{M}}={{10}\over{9}}\Big({{Q}\over{M}}\Big)^{2}\cdot\sqrt{\epsilon}\
\end{equation}
for the effective widths of the scalar clouds in the charged Reissner-Nordstr\"om black-hole spacetime (\ref{Eq4}). 

Substituting the analytically derived functional relation (\ref{Eq38}) into Eq. (\ref{Eq41}), 
one finds the dimensionless expression
\begin{equation}\label{Eq42}
{{\Delta r(Q/M,M\mu)}\over{M}}={{1}\over{{\bar\eta}^{{1\over4}}}}\cdot{{50}\over{9\sqrt{6\sqrt{15}}}}
\Big({{Q}\over{M}}\Big)^{4}
\Big({{{25Q^2-21M^2}\over{Q^2}}}\Big)^{1\over4}\cdot\sqrt{n+{1\over2}}\
\end{equation}
for the effective widths of the supported massive scalar field configurations, which in the large-mass regime (\ref{Eq14}) 
can be expressed in the form [see Eq. (\ref{Eq39})]
\begin{equation}\label{Eq43}
{{\Delta r(Q/M,M\mu)}\over{M}}={{Q}\over{M}}\cdot{{2^{5\over4}}\over{3}}
\Big({{{25Q^2-21M^2}\over{Q^2}}}\Big)^{1\over4}\cdot\sqrt{n+{1\over2}}\cdot{{1}\over{\sqrt{M\mu}}}\  .
\end{equation}
From Eq. (\ref{Eq43}) one learns that, in the dimensionless 
large-mass $M\mu\gg1$ regime, the supported massive scalar clouds in the 
charged Reissner-Nordstr\"om black-hole spacetime (\ref{Eq4}) can be made arbitrarily thin \cite{Noteemp}. 

\section{Penetration depth of the massive scalar field into the classically forbidden region}

In the present section we shall analyze the spatial behavior of the non-minimally coupled 
massive scalar field configurations outside the narrow radial interval (\ref{Eq40}) [see also (\ref{Eq43})] of 
the classically allowed region. 

It is important to stress the fact that the scalar field is not strictly zero outside the classically allowed 
region ({\ref{Eq40}). However, as we shall now demonstrate explicitly, the effective penetration depth of the 
scalar eigenfunction into the classically forbidden region becomes infinitesimally small 
in the large-mass $M\mu\gg1$ regime (\ref{Eq14}). 

In particular, according to the standard WKB analysis \cite{WKB1,WKB2,WKB3}, 
the radial function that characterizes the thin massive scalar field configurations is exponentially 
suppressed
\begin{equation}\label{Eq44}
\psi_{\text{WKB}}(r;M,Q,\mu,\bar\eta)\sim e^{-\int{\sqrt{V(y)}dy}}\  ,
\end{equation}
in the classically forbidden region outside the narrow radial interval (\ref{Eq40}). 
Interestingly, and most importantly for our analysis, one finds the relation [see Eqs. (\ref{Eq21}) and (\ref{Eq39})] 
\begin{equation}\label{Eq45}
\sqrt{V}\propto M\mu\gg1\
\end{equation}
in the large-mass regime (\ref{Eq14}). 
Thus, the effective penetration depth of the WKB scalar eigenfunction into 
the classically forbidden region [that is, into the region outside the classically allowed narrow radial 
interval (\ref{Eq40})] scales as $1/\mu$ and therefore becomes 
infinitesimally small in the large mass $M\mu\gg1$ regime (\ref{Eq14}) that we consider in the present paper. 

For example, for a Reissner-Nordstr\"om black hole with $Q/M=0.99$ supporting a non-minimally 
coupled massive scalar field with $M\mu=100$ one finds from Eqs. (\ref{Eq18}), (\ref{Eq21}), (\ref{Eq39}), 
and (\ref{Eq44}) the extremely small ratio $\psi(r=r_+)/\psi(r=r_{\text{peak}})\simeq1.2\times 10^{-219}$ 
[It is important to note that this dimensionless ratio becomes even smaller for larger field masses. For example, 
for a supported massive scalar field with $M\mu=1000$ one finds the characteristic 
small ratio $\psi(r=r_+)/\psi(r=r_{\text{peak}})\simeq8.7\times 10^{-2190}$].  

\section{Summary and Discussion}

Recent studies of the composed Einstein-Gauss-Bonnet-scalar field theory have revealed 
the physically intriguing fact that asymptotically flat black holes with spatially regular horizons 
can be spontaneously scalarized 
\cite{Sot5,Sot1,GB1,GB2,ChunHer,SotN,Hodsg1,Hodsg2,Hodca,Done,Herrecnum,BeCo,Brih,Hernn}.  

Motivated by this highly interesting observation, we have studied, using analytical techniques, 
the physical and mathematical properties of static matter 
configurations (linearized scalar clouds) which are made of {\it massive} scalar fields with a direct non-minimal coupling to the 
Gauss-Bonnet curvature invariant ${\cal G}$ of a central supporting charged 
Reissner-Nordstr\"om black hole. 

The main results derived in this paper and their physical implications are as follows:

(1) We have derived the charge-dependent upper bound (\ref{Eq27}) on the allowed proper masses of the 
non-minimally coupled scalar fields that can be supported in the curved spacetimes of charged 
Reissner-Nordstr\"om black holes. 

(2) It has been explicitly proved that, in the highly charged 
${{Q}/{M}}\geq ({{Q}/{M}})_{\text{crit}}={{\sqrt{21}}\over{5}}$ regime [see Eq. (\ref{Eq23})], 
the addition of a mass term to the
supported scalar fields allows the existence of {\it thin} static 
scalar shells that are non-minimally coupled to the Gauss-Bonnet curvature invariant and 
hover a {\it finite} proper distance above the horizon of the central charged black hole. 
In particular, the supported thin shells are characterized by the dimensionless relation [see Eq. (\ref{Eq22})] 
\begin{equation}\label{Eq46}
{{r_{\text{peak}}}\over{M}}={5\over3}\Big({{Q}\over{M}}\Big)^2>{{r_+}\over{M}}
\ \ \ \ \text{for}\ \ \ \ {{Q}\over{M}}>\Big({{Q}\over{M}}\Big)_{\text{crit}}={{\sqrt{21}}\over{5}}\  .
\end{equation}

(3) It has been shown that the supported scalar clouds 
are characterized by the effective dimensionless widths \cite{Notesubn01}
\begin{equation}\label{Eq47}
{{\Delta r(Q/M,M\mu)}\over{M}}={{Q}\over{M}}\cdot{{2^{3\over4}}\over{3}}
\Big({{{25Q^2-21M^2}\over{Q^2}}}\Big)^{1\over4}\cdot{{1}\over{\sqrt{M\mu}}}\  .
\end{equation}
Intriguingly, the analytically derived functional expression (\ref{Eq47}) implies 
that the supported scalar configurations, which are made 
of non-minimally coupled massive scalar fields, can be made arbitrarily thin 
in the dimensionless large-mass $M\mu\gg1$ regime.  

(4) Using a WKB analysis, we have derived the remarkably compact analytical resonance formula \cite{Notesubn02}
\begin{equation}\label{Eq48}
\sqrt{\bar\eta_0(Q/M,M\mu)}={{625}\over{54\sqrt{30}}}\Big({{Q}\over{M}}\Big)^{6}\cdot M\mu+
{{25}\over{12\sqrt{15}}}\Big({{Q}\over{M}}\Big)^{4}\sqrt{{{25Q^2-21M^2}\over{Q^2}}}\
\end{equation}
for the critical existence-line that characterizes 
the composed Reissner-Nordstr\"om-black-hole-nonminimally-coupled-linearized-massive-scalar-field cloudy configurations 
in the dimensionless large-mass (large-coupling) $M\mu\gg1$ regime. 

Finally, it is worth emphasizing the fact that the analytically derived 
critical existence-line (\ref{Eq48}) marks, in the dimensionless large-mass regime (\ref{Eq14}), the sharp 
boundary between bald black-hole solutions of the Einstein-Maxwell-Gauss-Bonnet-massive-scalar field theory (\ref{Eq1}) 
and hairy (scalarized) black-hole spacetimes that characterize the composed physical system. 

\bigskip
\noindent
{\bf ACKNOWLEDGMENTS}
\bigskip

This research is supported by the Carmel Science Foundation. I would
like to thank Yael Oren, Arbel M. Ongo, Ayelet B. Lata, and Alona B.
Tea for helpful discussions.



\begin{thebibliography}{99}
\bibitem{NoteQb} One can assume, without loss of generality, that the electric charge of the central supporting 
black hole is characterized by the relation $Q\geq0$.
\bibitem{Notemuu} Here the physical field parameter $\mu$ stands for $\mu/\hbar$. 
Hence, the mass parameter of the non-minimally coupled scalar field has the dimensions of $\text{length}^{-1}$.
\bibitem{Noteun} We shall use gravitational units in which $8\pi G=c=\hbar=1$.
\bibitem{Noteunir} It is interesting to stress the fact that the charge-dependent critical 
existence-line of the physical system, which marks the boundary between bald and hairy black-hole solutions 
of the field equations, is universal in the sense that it characterizes different Einstein-Maxwell-scalar field theories 
whose weak scalar field behavior is described by the leading-order functional expansion 
$f(\phi)\propto\phi^2\cdot[1+O(\phi^2)]$ \cite{Brih,Hernn}.
\bibitem{Noteetaa} The physical parameter $\eta$, which couples the supported massive scalar field $\phi$ to the 
Gauss-Bonnet curvature invariant ${\cal G}$, has the dimensions of length$^2$.
\bibitem{Notebl} We use here the Schwarzschild spacetime coordinates $(t,r,\theta,\phi)$.
\bibitem{Noteomt} For brevity, we shall henceforth omit the angular harmonic indexes $\{l,m\}$ of the 
supported massive scalar fields.
\bibitem{Noteppmm} Note that the new radial coordinate belongs to the infinite regime $y\in[-\infty,\infty]$.
\bibitem{Notenhm} Note that the functional expression (\ref{Eq21}) is valid in the $r=O(M)$ regime 
[see Eqs. (\ref{Eq14}), (\ref{Eq15}), and (\ref{Eq22})].
\bibitem{Notensz} It is important to emphasize that we shall analyze in the present paper scalar field configurations which are 
not strictly zero at the black-hole horizon. In particular, we shall consider supported 
massive scalar fields which, in the large-mass regime (\ref{Eq14}), are exponentially 
suppressed outside a classically allowed narrow radial interval [see Eqs. (\ref{Eq40}) and (\ref{Eq43}) below] with a penetration 
depth into a classically forbidden region that scales as $1/\mu$ and therefore becomes infinitesimally small 
in the large-mass regime $M\mu\gg1$.
\bibitem{NoteQQc} Here we have used the fact that, in the dimensionless sub-critical regime 
${{Q}/{M}}\leq ({{Q}/{M}})_{\text{crit}}$, the Gauss-Bonnet term (\ref{Eq20}) 
is a monotonically decreasing function in the exterior $r\geq r_+$ region of the black-hole spacetime, 
which implies the relation $V_{\text{GB}}(r\geq r_+;M,Q)\leq V_{\text{GB}}(r_+;M,Q)={{48M^4}\over{r^6_+}}-{{96M^3Q^2}\over{r^7_+}}+{{40M^2Q^4}\over{r^8_+}}$ 
in the regime ${{Q}/{M}}\leq \sqrt{21}/5$.
\bibitem{NoteMQ} Here we have used the inequality ${{2M}/{r^3}}-{{2Q^2}/{r^4}}\geq0$ which 
characterizes the exterior $r\geq r_+$ region of the black-hole spacetime (\ref{Eq4}).
\bibitem{Notexep} Here we have used the assumptions $|x|\ll 21(Q/M)^2-25$ and $\epsilon\ll [21(Q/M)^2-25]^2$. 
Taking cognizance of Eqs. (\ref{Eq36}), (\ref{Eq37}), and (\ref{Eq38}), one finds that these assumptions correspond to the large-mass 
requirement $M\mu\gg [21(Q/M)^2-25]^{-3/2}$.
\bibitem{Noteintegral} Here we have used the simple integral relation
$\int_{-1}^{1}dz \sqrt{1-z^2}={{\pi}/{2}}$.
\bibitem{Noteemp} It should be emphasized that the massive scalar field is exponentially 
suppressed in the classically forbidden region outside the narrow radial interval (\ref{Eq43}).
\bibitem{Notesubn01} Here we have substituted $n=0$ in Eq. (\ref{Eq43}) for the fundamental mode of the 
composed black-hole-massive-scalar-field system.
\bibitem{Notesubn02} Here we have substituted $n=0$ in the analytically derived resonance spectrum (\ref{Eq39}) 
for the fundamental mode of the 
composed black-hole-massive-scalar-field system.
\end{thebibliography}
\end{document}